\documentclass{iopart}

\usepackage{graphicx}

\begin{document}

\title[Simulations of Quantum XXZ Models on Two-Dimensional Frustrated Lattices]
{Simulations of Quantum XXZ Models on Two-Dimensional Frustrated Lattices}

\author{Roger G. Melko}
\address{
Materials Science and Technology Division,
Oak Ridge National Laboratory,
Oak Ridge, Tennessee 37831 USA
}

\date{\today}

\begin{abstract}
We report recent progress in the study of a particular class of spin 1/2 XXZ model on two-dimensional lattices with frustrated diagonal and unfrustrated off-diagonal interactions.  Quantum Monte Carlo simulations can be constructed without a sign problem, however they require non-trivial algorithmic advances in order to combat freezing tendencies.   We discuss results obtained using these techniques, in particular the discovery of unusual bulk quantum phases, studies of quantum criticality, and the continuing search for exotic physics in these models.
\end{abstract}




\section{Introduction}

Although widely believed to harbor a fertile playground for interesting and exotic physics, models of
frustrated quantum many-body systems are notoriously difficult to handle theoretically.  Numerical techniques offer some of the most promising and powerful tools in this regard. The largest-scale unbiased method for solving the quantum many-body problem is the quantum Monte Carlo (QMC) technique.  Unfortunately, QMC simulations suffer from the infamous ``sign problem'', which precludes arguably some of the most interesting physical cases from being accessed: frustrated quantum antiferromagnets and fermion systems at low temperatures.  Apart from the sign problem, QMC simulations may suffer from sever ``freezing'', or loss of ergodicity, particularly in models with competing interaction terms or a large degree of geometric frustration.

Over the past several years, a particular class of frustrated quantum spin/boson system has been made accessible to QMC simulations due to advanced algorithmic techniques designed to combat this freezing phenomenon.  This class of model is the spin 1/2 XXZ model, which can be formulated without a sign problem for ferromagnetic $J_{\perp}$ and antiferromagnetic $J_z$. 
The Hamiltonian is
\begin{equation}
H = -J_{\perp}\sum_{\langle i,j \rangle} ( S^x_iS^x_j + S^y_iS^y_j ) +   J_z\sum_{\langle i,j \rangle }S^z_i S^z_j - h\sum_i S^z,
\label{XXZham}
\end{equation}
where we restrict the nearest-neighbor ({\it nn}) exchange $J_{\perp} > 0$, and frustration arises from $J^z>0$.  
This model is equivalent to hard-core bosons with {\it nn} hopping amplitude $t=2J_{\perp}$ and repulsive interaction $V=J_z$, where making the usual substitution  ($S^z_i = n_i - 1/2$, $S^+_i = b^{\dagger}_i$ etc.) results in the mapping.

The simplest XXZ spin Hamiltonians on two-dimensional (2D) frustrated lattices have recently revealed fascinating non-trivial quantum phases.  The two most exciting examples are a stable supersolid phase in the {\it nn} triangular lattice XXZ model \cite{sstri1,sstri2,sstri3}, and a spin singlet or valence-bond solid (VBS) phase in the kagome lattice \cite{KagomeVBS,KagomeDamle} model.
In addition, the simplest extension of this class of Hamiltonian on the triangular lattice, obtained by adding a frustrating diagonal next-nearest neighbor ({\it nnn}) term $J_z^{\prime} \sum_{\langle \langle i,j \rangle \rangle} S^z_i S^z_j$, has recently revealed the presence of another stable supersolid phase that exhibits stripe-like symmetry \cite{NNNTri}.  Both the {\it nn} kagome and {\it nnn} triangular lattice models have also offered valuable testing grounds in the search for unconventional (non-Landau) quantum critical behavior \cite{KagomeVBS,NNNTri,KagomeDVT,theBurkonian}.

In the next section of this paper, we review the most recent algorithmic advances that make QMC simulations of this class of Hamiltonian possible on 2D frustrated lattices.  We then present a selection of results on the models mentioned above, with a focus on relevance to interesting bulk quantum behavior, searches for exotic spin liquids, and potential examples
of unconventional quantum criticality.

\section{Stochastic Series Expansion QMC}

State of the art QMC techniques for the study of quantum Hamiltonians in the spin 1/2 XXZ class 
employ a variety of updating schemes to sample world-line configurations in a continuous imaginary-time framework \cite{KawHarRev}.  These techniques are free of the systematic error associated with a Trotter-type discretization of imaginary time, and give full access to finite-temperature estimators in a grand-canonical ensemble.  The QMC variant employed in the present paper is the {\it stochastic series expansion} (SSE) technique, pioneered by Sandvik \cite{sse1,sse3,sse4}. 

In the SSE formalism, one is interested in sampling operator sequences (particle trajectories or world-lines) in the $d+1$ dimensional simulation cell using a Metropolis rejection scheme, with transition probabilities derived from the partition function.
An important element of the technique, which makes sampling of off-diagonal operators feasible, is the efficiency gained with implementation of advanced global or ``loop'' moves in the QMC.
In the SSE framework, 
the {\it directed-loop} equations are used to construct transition probabilities that eliminate loop ``bounces'' or ``back-tracking'' (i.e. the tendency of a loop to trace back over its own path).  The elimination of bounces has been shown to be a critical element in the construction of efficient and ergodic codes for the XXZ class of model.
The reader is referred to Reference \cite{DIRloop} for a comprehensive introduction to sampling schemes in the SSE QMC, including a detailed explanation of the directed-loop algorithm.

The starting point for any SSE simulation scheme is the construction of
an operator list, the elements of which are sampled using the various updating algorithms.  
This is done by first writing the Hamiltonian as a sum of elementary interactions,
\begin{equation}
H = - \sum\limits_{t}\sum\limits_{a} H_{t,a},
\label{hsum}
\end{equation}
where in a chosen basis $\{ |\alpha \rangle \}$ (e.g.~the standard $S^z$ basis) the operators satisfy
\begin{equation}
H_{t,a}|\alpha \rangle \sim |\alpha^\prime \rangle ,
\end{equation}
and $|\alpha \rangle$ and $|\alpha^\prime \rangle$ are both basis states.
The index $t$ refers to the operator types (various kinetic and
potential terms), while $a$ is the lattice units over which the interactions are
summed.  
The challenge at this stage is to devise an intelligent way to decompose the full Hamiltonian ($\ref{XXZham}$) into the individual elements in \Eref{hsum}, over which sampling takes place.
In the original SSE algorithm \cite{DIRloop}, the basic lattice units are {\it bonds} connecting two single sites.  In the next section we discuss an alternative scheme for performing this Hamiltonian decomposition, which is used in simulations of frustrated XXZ models, and explain why it results in a large increase in sampling efficiency in some parameter regimes.

\subsection{Triangular plaquette decomposition}
\label{TplaqSS}

\begin{figure}[floatfix]
\begin{center}
\includegraphics[width=3.5in]{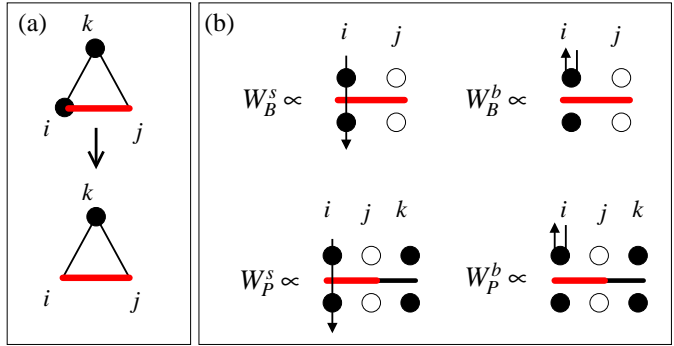}
\caption{A transition between two ``minimally'' frustrated plaquettes is illustrated in (a).  
In (b), a loop move that attempts to remove the lower left particle in the traditional bond-based algorithm (top) is suppressed,
since the probability of the loop bouncing out along the entrance leg is much greater than the probability that it will pass through the vertex ($W^b_B >> W^s_B$).  
In contrast, in the plaquette-based algorithm (bottom) the 
weight of both processes is equal: $W^s_P = W^b_P = W^{\rm mf}$ (see \Eref{Wmf}).
\label{SSEplaq}}
\end{center}
\end{figure}

Any Metropolis Monte Carlo algorithm that attempts to simulate certain parameter regimes (in our case large $J_z$ or $J_z^{\prime}$) of models on 2D frustrated lattices (e.g. triangular, kagome, or other lattices composed of triangular ``plaquettes'') will tend to encounter difficulty with freezing or loss of ergodicity.
In this regime, the system is strongly influenced by the classical ground state, which can be described as a highly-degenerate manifold of minimally-frustrated triangular plaquettes (each with two-up and one-down spin, or vice versa).  
If the classical system is perturbed by a small $J_{\perp}$, this degeneracy is slightly lifted.
Simulation freezing occurs due to the inability of the updating algorithms to traverse the nearly-degenerate basis configurations that result.  Large energy barriers exist between these configurations: the creation of a ``defect'' plaquette (with more than one frustrated bond) is strongly suppressed, and the type of global cluster move traditionally used to combat this in classical simulations is extremely difficult to implement in a QMC framework.  

In the bond-based SSE operator sampling scheme \cite{DIRloop}, the existence of this near-degeneracy in ground state basis configurations results in a type of ``local'' energy barrier in the operator-loop update.
As discussed in detail below, this energy barrier suppresses the transition probability between different minimally-frustrated local lattices plaquettes, resulting in a loss of efficiency in the algorithm.
This difficulty can be overcome by a removal of the local energy barrier, done by constructing the decomposition of the Hamiltonian as a list of three-site triangular plaquettes \cite{Tplaq,SergeiPC}.
For example, a diagonal term $H_{1,a}$ for a frustrated {\it nn} interaction would be written (c.f.~Ref.~\cite{DIRloop}),
\begin{equation}
H_{1,a} = C - J_z \left[{S_i^z S_j^z+S_j^z S_k^z + S_k^z S_i^z }\right] -\frac{h}{z} \left[{S_i^z+S_j^z+S_k^z}\right],
\label{Hdiag1}
\end{equation}
where the site indices $i$, $j$ and $k$ belonging to the plaquette $a$ are illustrated in Figure~\ref{SSEplaq}a, and $C$ is some constant necessary to keep the corresponding transition probabilities positive definite (and hence avoid the sign problem).  The diagonal term for triangular-lattice {\it nnn} interactions is identical, save for the substitution of $J_z^{\prime}$ for $J_z$.  Note that the scaling of the $h$ field is dependent on the coordination of the lattice $z$, which may include {\it nn} or {\it nnn} bonds depending on the specific model studied.  In \Eref{Hdiag1} the index 1 allows for other types of interactions, for example the {\it nn} quantum exchange (or hopping) which is defined analogous to Ref.~\cite{DIRloop}.

Transition probabilities used in the Metropolis Monte Carlo sampling are derived from the partition function: in the SSE representation it can be written \cite{sse1,DIRloop},
\begin{equation}
Z = \sum\limits_\alpha \sum_{n=0}^{\infty}  \sum_{S_n} \frac{(- \beta)^n}{n!}
    \left \langle \alpha  \left | \prod_{i=1}^n H_{t_i,a_i}
    \right | \alpha \right \rangle ,
\label{Zpart}
\end{equation}
where $\beta$ is the inverse temperature $1/T$,
and $S_n$ is the operator-index sequence
$S_n = [t_1,a_1],[t_2,a_2], \ldots ,[t_n,a_n]$ over which sampling occurs.  
The operator product can be viewed as propagating the state $| \alpha \rangle$ in the (+1) expansion dimension: $| \alpha(\tau) \rangle \sim \prod_{i=1}^{\tau} H_{t_i,a_i}|\alpha \rangle $, so that the weight factor associated with \Eref{Zpart} is
\begin{equation}
W = \frac{\beta^n}{n!} \prod_{\tau=1}^{n}  W(\tau), \label{Weight}
\end{equation}
defined in terms of the matrix element $W(\tau)=\langle \alpha(\tau) | H_{t_{\tau},a_{\tau}} | \alpha(\tau-1) \rangle$.

Transitions probabilities for the various updates are easily calculated from the matrix elements in \Eref{Weight}.  In the case of the {\it nn} triangular plaquette-based SSE algorithm, there are two diagonal weights at $h=0$ -- a fully-frustrated (ff) and a minimally-frustrated (mf):
\begin{eqnarray}
W^{\rm ff} &=& \langle \uparrow \uparrow \uparrow | H_{1,a} | \uparrow \uparrow \uparrow \rangle
= \langle \downarrow \downarrow \downarrow | H_{1,a} | \downarrow \downarrow \downarrow \rangle
=C-3J_z/4 \\
W^{\rm mf} &=& \langle \uparrow \downarrow \uparrow | H_{1,a} | \uparrow \downarrow \uparrow \rangle
=\langle \downarrow \uparrow \downarrow | H_{1,a} | \downarrow \uparrow \downarrow \rangle
=C+J_z/4, \label{Wmf}
\end{eqnarray}
where $C \geq 3J_z/4$ is required to keep the weights positive-definite.
In this plaquette-based decomposition, six of eight diagonal matrix elements are minimally frustrated (one ferromagnetic bond), as opposed to two of four being unfrustrated (ferromagnetic) in the bond-based algorithm \cite{DIRloop}.  This fact may be expected to increase the efficiency of the diagonal update
 in choosing a minimally-frustrated plaquette; however, a larger increase in efficiency is expected in the operator-loop updates, which as discussed above transform one type of matrix element (or ``vertex'') into another,  and sample both diagonal and off-diagonal operators.

In the operator-loop, the simulation cell is divided into vertices representing the propagation of the basis state by the operators in $H$.    In the plaquette-based SSE, each vertex has six ``legs'' representing the $S^z$ basis elements before and after a given operator acts.
The loop ``head'' enters a vertex at a given leg, and the probability that it exits at another leg is determined by a weight factor calculated from the partition function discussed above.
Many different solutions for the exit probabilities are possible.  In the simplest (heat-bath) solution, the probability of selecting an exit leg is proportional to the weight of the resulting matrix element \cite{DIRloop}.  Thus, as illustrated in Fig.~\ref{SSEplaq}, 
the probability of a loop exiting such that a different minimally-frustrated plaquette is sampled is at least equal to the probability of bouncing out along the entrance leg.  In contrast, in the original bond implementation of the SSE algorithm, the loop is much more likely to follow a bounce path to keep the bond antiferromagnetic, or to sample an off-diagonal vertex proportional to the exchange $J_{\perp}$ (which however is assumed to be relatively small).
Obviously, even though
this simple demonstration 
illustrates the concept that the plaquette-based algorithm promotes an increase in relevant sampling, it is important to remember that plaquette-based sampling {\it alone} is not typically sufficient to produce an efficient simulation. 
Additional algorithmic advances are also necessary in this regard,
in particular the implementation of the directed-loop solutions \cite{DIRloop} to the weights discussed, in order to further reduce or eliminate the bounce probability wherever possible.
 
\begin{figure}[ht]
\begin{center}
\includegraphics[height=4.8cm]{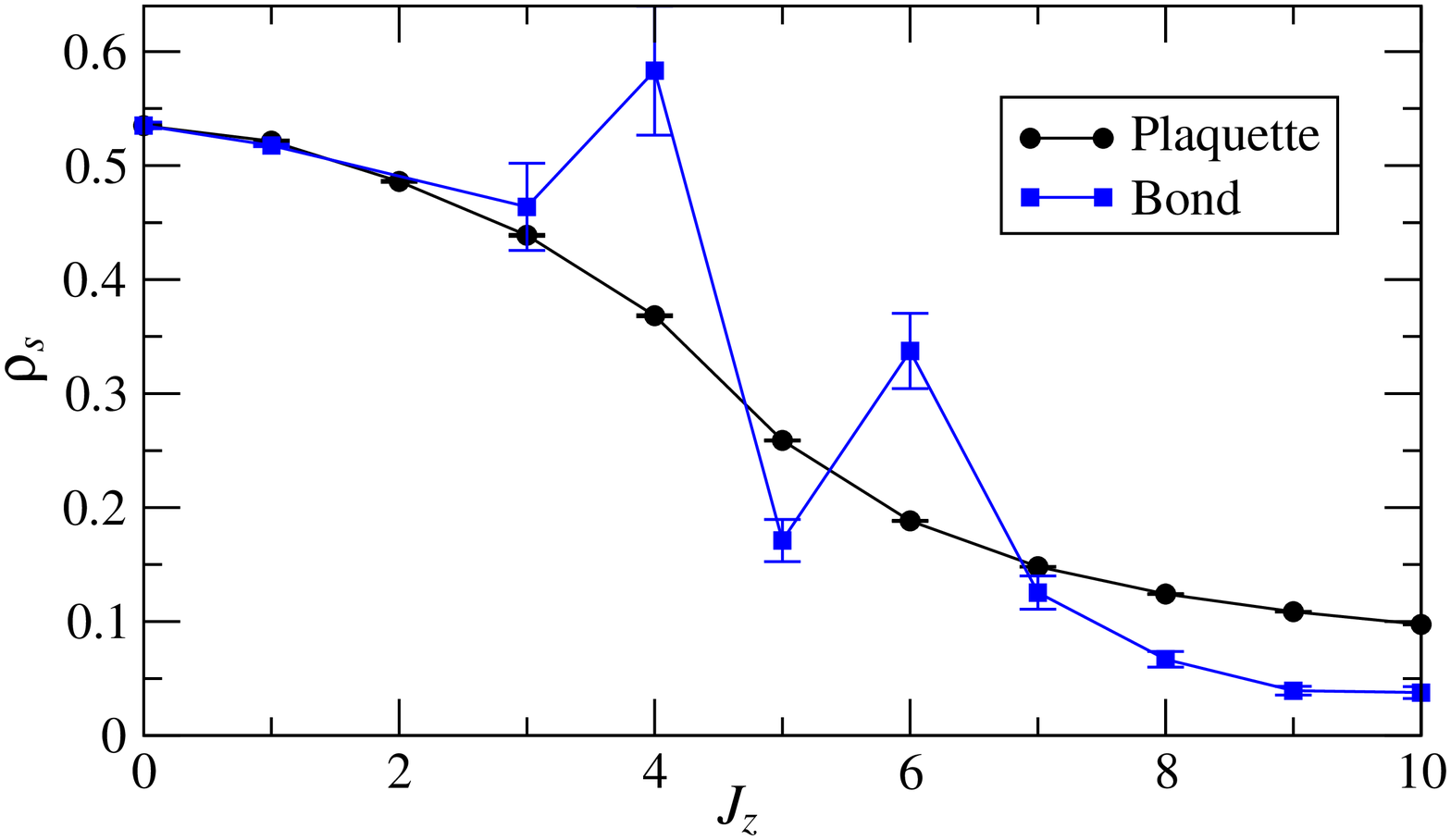}
\caption{
Spin stiffness $\rho_s$ for the {\it nn} XXZ model on a $9 \times 9$ triangular lattice at $J_{\perp}=1$ and $T=0.1$.  Both the plaquette version and the bond version of the SSE algorithm were ran with $10^6$ QMC production steps.   
}
\label{rhoPB}
\end{center}
\end{figure}

An example of the statistical improvement for data obtained with this modified QMC is illustrated in Figure~\ref{rhoPB}, where the spin stiffness (superfluid density) is illustrated for both the bond- and plaquette- based algorithms.  There, in addition to the obvious decrease in noise across a phase boundary (which occurs at $J_z \approx 4.5$) with the plaquette algorithm, it is clear that the bond-based algorithm is in danger of severely under-estimating the value of $\rho_s$ deep in the large-$J_z$ phase.  

Finally, one should note that although the plaquette-based algorithm is successful in eliminating an algorithmic energy barrier associated with the update of a local region of minimally frustrated plaquettes,
it does not remove the {\it global} energy barriers associated with creating defects in the entire near-degenerate manifold of $S^z$ configurations (discussed at the beginning of this section).
Therefore one can ultimately expect any QMC simulation to lose ergodicity and experience freezing for sufficiently large $J_z$ (or $J_z^{\prime}$), even with the plaquette implementation.  
The most sever freezing observed recently on this class of model occurred in the {\it nnn} triangular lattice case \cite{NNNTri}, where signs of ergodicity loss were experienced for $J_z^{\prime} \ge 5$.  The {\it nn} kagome model had the least sever freezing, with simulations remaining ergodic to $J_z=50$ or higher even at low temperatures \cite{KagomeVBS}.  
In cases where the exploration of ground-state physics is still hampered regardless of the advances outlined in this section, 
other algorithmic improvements may be necessary to combat this global freezing phenomenon.  We discuss one example below.

\subsection{Quantum Parallel Tempering}

The parallel tempering (PT) algorithm is a powerful extension to the usual single-Markov chain Metropolis QMC algorithm \cite{LiuMC}.  Thermal parallel-tempering has been used extensively for simulations of frustrated systems, spin glasses, polymers, and an abundance of other classical systems.  The method is straightforwardly generalized for use on quantum systems, and a scheme for implementing it in the SSE framework was first proposed by Sengupta {\it al.} \cite{PTsse}.

The PT scheme is employed to combat algorithmic freezing -- in our case this freezing is due to the large energy barriers that occur in the near-degenerate manifold of basis states at low temperature and large diagonal interaction.  Assuming that we can access both the frozen and unfrozen regimes of the model, the basic framework of the PT algorithm is:
\begin{enumerate}
\item{} Run a number of replicas $N_{x}$ of the simulation in parallel, using a set of ``adjacent'' parameter values  $\{ x \}$ (e.g. temperature), some of which are in the frozen regime, some outside of it.
\item{} After some number of conventional QMC steps, attempt to ``swap'' the configurations of neighboring parameter bins.
\end{enumerate}
In this way, frozen replicas can benefit from the superior efficiency of the unfrozen replicas, which are able to share their sampled configurations through the PT swap.  The effective number of relevant configurations that the replicas in the frozen regime can sample is therefore vastly increased, resulting in a significant  ergodicity improvement.

The swapping step is standard and is described elsewhere \cite{LiuMC}.  The most straightforward and commonly used tempering schemes vary the temperature in adjacent QMC simulations, although varying other quantum parameters is also possible \cite{PTsse}.  In a SSE PT algorithm where replicas differ in temperature, the step (ii) involves proposing swaps of configurations (the operator-index sequence $S_n$ and basis state $|\alpha \rangle$) between {\it adjacent} temperature bins, $\beta_i \leftrightarrow \beta_{i+1}$, starting with the highest temperature and ending with the lowest.  The swap moves are accepted or rejected based on a Metropolis algorithm, with a probability derived from \Eref{Weight}
that is the product of weights after the swap over the product before:
\begin{equation}
P_{\rm swap}(i,i+1) = {\rm min}\left[{1, \left({ \frac{\beta_{i+1}}{\beta_i} }\right)^{n_i-n_{i+1}}  }\right].
\end{equation}

\begin{figure}[floatfix]
\begin{center}
\includegraphics[width=3.3in]{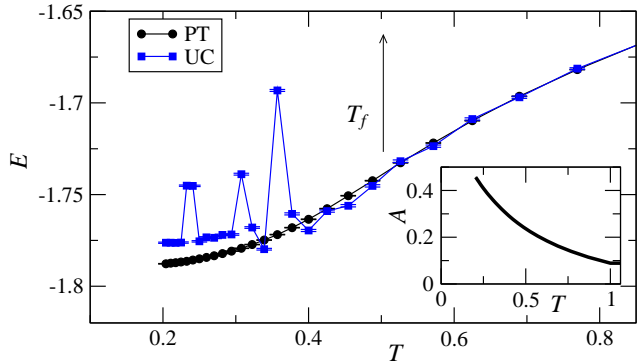}
\caption{
System energy ($E$) calculated using identical {\it bond}-based QMC simulations on a $6 \times 6$ triangular lattice with $J_z=6$ and $J_z^{\prime}=0$, with (PT) and without (uncorrelated -- UC) a PT swap step between temperature bins.  The approximate freezing temperature $T_f$ is evident.  The inset is the acceptance rate ($A$) of the PT swap, which depends highly on the choice of the set $\{ \beta_i \}$.
\label{PTfig}}
\end{center}
\end{figure}

Perhaps the most difficult practical aspect involved in implementing a PT algorithm is deciding on a set of $\beta_i$ values that gives a useful acceptance rate for the PT swaps.   A set with a constant temperature spacing $\delta T$ will give too small an acceptance rate in the low-$T$ region.  We employ the next simplest option in simulations of the {\it nn} triangular lattice XXZ model -- constructing the set $\{ \beta_i \}$ to have constant $\delta \beta$ spacing.  Obviously more sophistocated choices are also available.
Results from some PT simulations are illustrated in Figure~\ref{PTfig}.  As is clear in this example, the PT scheme removes most evidence for freezing below $T_f$ and improves the lowest-temperature equilibriation (which we are most interested in presumably).  
It also obviously gives access to the full temperature information in one run, although in cases where one is not interested in the finite-$T$ physics, the increased CPU time involved in running all $T$ bins must be weighed against the CPU time required to equilibriate one low-$T$ bin alone.
However, as illustrated in Figure~\ref{PTfig}, it is apparent from the size of the error bars that the data from the non-PT (uncorrelated) simulation is essentially ``stuck'' in their respective configurations regardless of convergence to the ground state, while the PT data is much better converged.  Thus, in this case the PT algorithm is a rare example of a non-trivial parallelization improvement in a QMC simulation.

\section{Ground state phase diagrams for several frustrated XXZ models}

\begin{figure}[ht]
\begin{center}
\includegraphics[height=4.5cm]{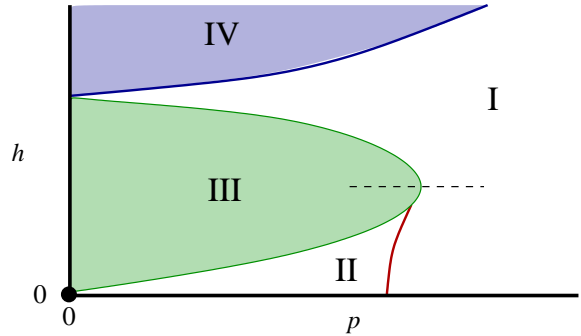}
\caption{
Schematic ground-state phase diagram for models with one frustrating parameter $p=J_{\perp}/J_z$ or $J_{\perp}/J_z^{\prime}$ (see text).  The phase diagram is particle-hole symmetric about the $p$-axis.
  Studies of quantum phase transitions employed in the search for exotic non-Landau critical points have been done between phases I and III, typically along a trajectory that matches the magnetization (filling) on both sides of the transition (e.g.~dashed line).
}
\label{PhaseD}
\end{center}
\end{figure}

Using some combination or variation of the algorithmic advances outlined above, much progress has been made over the past several years in illuminating the physics of models in the class of \Eref{XXZham} -- spin 1/2 XXZ systems on 2D frustrated lattices.
The simplest Hamiltonians in this class possess ground state phase diagrams with some common qualitative features.  With a single frustrating parameter ($J_z$ or $J_z^{\prime}$) they include several phases, the dominant of which is an in-plane ferromagnet ($\langle S^x \rangle >0$ at $T=0$) with zero magnetization, $m \equiv \langle S^z \rangle=0$, but characterized by a finite spin stiffness $\rho_s$ (or superfluid density) -- see region I of Figure~\ref{PhaseD}. In addition to the trivial fully-polarized ferromagnetic phases at large $|h|$ (region IV), ``lobes'' of $m=\pm 1/6$ (or 2/3 and 1/3 filled Mott insulating) states exist (region III)\cite{MurthyMFT}.  Although occupying qualitatively similar positions on the phase diagram, the exact nature of these $m=\pm 1/6$ states is found to be drastically different for the two simplest cases of the {\it nn} triangular and kagome models studied recently, which we discuss below.   
Finally, the phase adjacent to these lobes -- region II -- is typically found to have a non-vanishing spin stiffness (hence off-diagonal long-range order (ODLRO)), but in special cases can also acquire coexisting diagonal long-range order, as discussed below. 

It is an interesting perspective to consider these various phases as developing by ``perturbing'' the respective classical frustrated Ising models (the $(0,0)$ origin of Figure~\ref{PhaseD}) with the addition of the off-diagonal $p$ (defined as  $J_{\perp}/J_z$ or $J_{\perp}/J_z^{\prime}$).  In the two simplest cases of {\it nn} interactions on the triangular and kagome lattices, the classical phases at $(0,0)$ are essentially unique, with different values for the extensive entropy arising from ground-state degeneracies.
In the triangular lattice, the extensive entropy per spin is $S/N=0.323$, while in the kagome it is $S/N=0.502$ \cite{IsingFrustMoes} (c.f.~a maximum value for the disordered Ising system of $S/N=\ln(2)=0.693$).  In the two cases studied where a classical Ising point with $S/N=0.323$ was perturbed by 
$p$ (see Figure~\ref{PhaseD}), a thermodynamically stable phase with both diagonal and off-diagonal long-range order is found \cite{sstri1,sstri2,sstri3,NNNTri}.  This phase is referred to as a supersolid phase in the boson language. In contrast, at the $(0,0)$ point on the {\it nn} kagome lattice phase diagram where $S/N=0.502$, perturbation by $p$ produces a phase with only uniform ODLRO.
In both of these bulk quantum phases, the ground state entropy of the classically degenerate manifold is lifted by a small perturbation in $J_{\perp}$.

In the {\it nn} triangular lattice, perturbation of the classical Ising $(0,0)$ point by a magnetic field $h$ is another mechanism by which to lift the classical ground-state degeneracy, producing a long-range ordered state with ordering wavevectors at $(4 \pi /3,0)$ and symmetry-related directions (region III in Figure~\ref{PhaseD}).  In contrast, at the kagome lattice Ising point, application of $h$ only {\it partially} lifts the extensive entropy, producing a disordered state with a reduced value of $S/N=0.108$ \cite{IsingFrustMoes}.  Qualitatively, this state persists along the $h$-axis of our schematic phase diagram, from the origin to the region III/IV phase boundary.  Interestingly, upon being perturbed by $J_{\perp}$, this phase acquires long-range order in both diagonal ($S^z_i$) and {\it bond} (i.e.~$S^+_iS^-_j + {\rm h.c.}$) correlations, but no ODLRO and zero spin stiffness \cite{KagomeVBS}.  This phase is a VBS, 
characterized by the formation of spin singlets around a bond (or series of bonds).  The VBS corresponds to only {\it partial} boson delocalization, and gives no winding number signature in the QMC, and hence zero spin stiffness.  Finally, in the recently studied {\it nnn} ($J_z^{\prime}>0$ and $J^z=0$) XXZ model on the triangular lattice, perturbation of the classical $(0,0)$ point by $h$ removes the extensive entropy of the system, but the resulting $m=1/6$ state retains a two-fold degeneracy associated with two equal-energy ground-states of different broken symmetry \cite{NNNTri}.  Upon further perturbation by $J_{\perp}$, the system selects one unique ground state -- a period-three striped Mott insulator that breaks rotational symmetry.

Far away from the small $p$ limit, the in-plane ferromagnet (superfluid) phase begins to dominate
(Figure.~\ref{PhaseD} region I), offering a unique opportunity to study interesting examples of quantum phase transitions between the diagonal and off-diagonal long-range ordered phases.  Particular attention has been paid recently to the region I/III phase boundary in the {\it nn} kagome and {\it nnn} triangular lattice XXZ models, where input from phenomenological field theories have suggested the possibility of exotic ``deconfined'' quantum criticality \cite{DQCP1} at points of equal magnetization \cite{KagomeDVT,theBurkonian}.  Results on the {\it nnn} triangular model have shown that the first initial requirement of the zero-temperature phase transition being continuous is clearly not realized \cite{NNNTri}, while on the {\it nn} kagome model, extremely large lattice sizes (over $10^4$ spins) are required to see very weakly first-order behavior \cite{KagomeVBS}, also precluding the existence of unconventional quantum criticality.

\section{Discussion}

For an alternative perspective, consider the above results in the boson language.
The superfluid phase corresponds to maximum disorder in the diagonal correlations of the system; the superfluid has only ODLRO.  Bosons are delocalized over the entire system, resulting in particle-trajectories that wind across the periodic boundaries in the $d+1$ dimensional system and hence a finite superfluid (spin) stiffness.  This full delocalization is also present in the supersolid, however in a homogeneous coexistence with diagonal long-range order in the density correlations.  This means that the bosons also have a tendency to localize on some lattice sites, and are not ``fully'' disordered in this sense.  In the VBS phase, bosons are strongly localized, but some may still dissociate with specific lattice sites and form delocalized or resonating bonds.  These ``strong'' bonds can extend over a moderate distance (six-bond delocalization is observed around a hexagonal ring in the {\it nn} kagome lattice model \cite{KagomeVBS}),
but are not fully delocalized over the lattice 
and do not result in a winding number and finite superfluid stiffness.  They do however give strong signatures in bond-bond correlation functions \cite{KagomeVBS,JKprl}.  Finally, the remaining Mott insulating states observed in these models correspond to a full localization of boson density, and no long-range correlations in boson hopping or bond strength are present.
 
\begin{figure}[ht]
\begin{center}
\includegraphics[height=2.1cm]{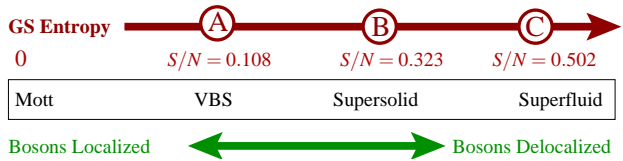}
\caption{
Values for the ground-state entropy in the classical minimally-frustrated manifold for the various models discussed in the text (top): (A) is the {\it nn} kagome with $h>0$; (B) is the {\it nn} and {\it nnn} triangular with $h=0$; (C) is the {\it nn} kagome with $h=0$.  As the entropy of the ground state increases, the degree of boson delocalization in the perturbed quantum system increases (bottom), leading to the various categories of bulk quantum phases listed (middle). 
}
\label{MottMelt}
\end{center}
\end{figure}

The interesting connection between boson localization in the bulk quantum phases that result from a ($p$) perturbation of some frustrated classical states is illustrated in Figure~\ref{MottMelt}.  A deeper understanding of the mechanisms which lead to the various degrees of boson delocalization require a proper handling of the small hopping using perturbation theory; however the underlying source of the allowed hopping processes depends on the connectivity of the lattice and therefore ultimately stems from the same source as the classical frustration.  In addition, one wonders whether extremely large geometric frustration may result in a ``total'' disordering of the bosons in both diagonal and off-diagonal correlations, resulting in a resonating valence bond (spin liquid) phase in this class of model.  Such a prospect seems unlikely -- away perhaps from special deconfined quantum critical points \cite{DQCP1} -- since the basic hopping parameter (the ferromagnetic $J_{\perp}$ or $t$) essentially remains unfrustrated.  This would appear to coincide with the observed tendency toward obtaining superfluid flow in these models, even for some regions on the phase diagram where the hopping is a perturbation.
The study of further examples of degenerate classical manifolds perturbed by quantum hopping processes would clearly shed more light on this discussion.

From a practical perspective, conclusions such as these are critically dependent on the development of efficient numerical algorithms that are able to simulate systems with a large number of spins/bosons.  Indeed, the basic indicators of bulk quantum phases, in particular the survival of Bragg peaks in the structure factor in the thermodynamic limit, may easily be missed if proper and rigorous finite-size scaling studies are not performed.  In addition, forays into studies of quantum criticality, e.g.~in effort to discern examples of exotic non-Landau quantum critical points, are possible on this class of model \cite{KagomeVBS,NNNTri}, however often require lattice sizes of thousands of sites in order to establish firm conclusions.  The ongoing effort to revamp and improve large-scale quantum Monte Carlo algorithms is hence a necessary requirement for the continuing success of numerical techniques in making important contributions to this field.

\section{Acknowledgments}
I extend heartfelt gratitude to A. Del Maestro, S. Isakov, A. Sandvik and D. Scalapino for aid and abetment.
Simulation data presented in this paper was obtained on the IBM p690 Power4 machine operated by the NCCS at ORNL.

\section*{References}

\bibliographystyle{iopart-num}
\bibliography{rmBiblio}

\end{document}